\documentclass[aip,jcp,reprint,showpacs,singlecolumn,superscriptaddress]{revtex4-1}
\usepackage{graphicx}
\usepackage{color}
\usepackage{amsmath}
\usepackage{comment}
\usepackage[export]{adjustbox}
\usepackage{placeins}
\newcommand{\be}{\begin{equation}}
\newcommand{\ee}{\end{equation}}
\newcommand{\bea}{\begin{eqnarray}}
\newcommand{\eea}{\end{eqnarray}}
\usepackage{dcolumn}
\usepackage{hyperref}
\usepackage{bm}
\usepackage{epsf}
\usepackage{subfigure}
\usepackage{epstopdf}%
\usepackage{amsfonts}

\begin{document}

\title{
Reconciling perturbative approaches
in phonon assisted transport junctions}

\author{Bijay Kumar Agarwalla and Dvira Segal}
\affiliation{Chemical Physics Theory Group, Department of Chemistry,
and Centre for Quantum Information and Quantum Control,
University of Toronto, 80 Saint George St., Toronto, Ontario, Canada M5S 3H6}

\date{\today}

\begin{abstract}
We present consistent results for molecular conduction using two central-complementary approaches:
the non-equilibrium Green's function technique and the quantum master equation method.
Our model describes electronic conduction in a donor-acceptor junction in which electron transfer
is coupled to nuclear motion, modeled by a harmonic vibrational mode.
This primary mode is further coupled to secondary phonon modes, a thermal bath.
Assuming weak electron-phonon coupling but arbitrary large
molecule-metal hybridization, we compute several non-equilibrium transport quantities:
the mean phonon number of the primary mode, charge current statistics.
We further present scaling relations for the cumulants valid in the large voltage regime.
Our analysis illustrates that the non-equilibrium Green's function technique
and the quantum master equation method can be worked out
consistently, when taking into account corresponding scattering processes.
\end{abstract}

\maketitle


\section{Introduction}
\label{intro}

Studies of charge transfer in single-molecule junctions \cite{TaoR,GalperinR,LathaR,RatnerR}
capture significant attention for various reasons:
(i) Molecular junctions offer a unique playground for exploring basic quantum effects in dynamics
e.g., interference between different transport pathways \cite{QI1,QI2}, and
(ii) molecules can serve as building blocks for electronic devices such as
molecular diodes \cite{Diode1, Diode2}, sensors, switches, thermo and opto-electronic devices
\cite{MEBook}. 

Electron-nuclei interactions are central to molecular electronic applications \cite{WangR}.
Nuclear motion in conducting molecules and the surroundings governs effects such as
local heating of the junction, which may lead to instabilities \cite{Lu}, and
incoherent tunnelling processes, responsible for the development
of ohmic conduction \cite{SegalET}. 
These effects can be captured within simple models:
The celebrated Anderson-Holstein model includes a
single electronic site embedded between metals,
further coupled to a single vibration, or a harmonic bath
\cite{galperinIN06, galperinIN07, Schmidt, Ora09, Levy1, Levy2,Levy3,Gogolin,thorwart,Thoss1,Simine14,Utsumi1,Joe2,Joe3}.
A different class of problems, relevant as well to photovoltaic devices \cite{Solar-review}, concerns
donor-acceptor type molecular systems, with two electronic sites coupled to molecular vibrations
\cite{Lu, SiminePCCP, INFPI3, Markussen, SimineTE,BijayDAPRB, BijayDABeil,ThossDA}.
Experimentally, inelastic electron tunnelling spectroscopy  can identify molecular
vibrations participating in the transport process \cite{IETS1, IETS2,IETSrev}.

Besides measurements of current-voltage characteristics \cite{TaoR,LathaR},
the distribution function of current fluctuations, or the
{\it full counting statistics} (FCS), can be received experimentally
by counting the number of tunneling electrons through a conductor within a given
time period \cite{FCS1,FCS2}.
FCS provides a complete picture of the transport problem
in the steady state regime 
with cumulants of current conveying information over interactions such
electron correlation effects, and the junction's geometry.

From a theoretical perspective, two eminent perturbative methods are routinely used to
investigate transport properties in molecular junctions:
the non-equilibrium Green's function (NEGF) technique
\cite{NEGF-review, NEGF-book1, NEGF-book2, NEGF-book3, Dhar-NEGF},
and the quantum master equation (QME) approach \cite{Jens,Mitra,Esposito-QME, Segal-QME,Ulrich06,wegeK,WangQME,Ulrich15}.
Both schemes are perturbative in nature, yet are developed along separate lines,
identifying different interaction Hamiltonians for the perturabtive expansion.

The NEGF theory is exact for bilinear metal-molecule Hamiltonians,
but it is perturbative in nonlinear interaction terms within the subsystem
(molecule), e.g., electron-vibration coupling or electron-electron interactions.
%
It hands over a formally exact expression for currents, the Meir-Wingreen (MW) formula \cite{Meir},
written  in terms of the so-called nonlinear self-energy.
The MW formula was recently generalized to accommodate
higher order fluctuations \cite{Gogolin-FCS, Bijay-Euro}.  
Furthermore, the NEGF approach can be used to reach analytical expressions for the cumulant generating
function (CGF), which directly provides the mean current and its fluctuations \cite{BijayDAPRB}.
QME approaches, in contrast, are typically developed in the
energy basis of the subsystem, thus
treating, in an exact manner, internal-molecular nonlinear interactions.
However, QME methods are typically  made
perturbative in the molecule-metal coupling \cite{Mitra, wegeK}, thus restricted
to handle only low-order processes in the tunnelling strength.
Equivalence between NEGF and QME in molecular transport problems has been demonstrated
only in the weak molecule-metal coupling limit in specific situations \cite{Peskin,Galperin15}.

In this paper, we  show that NEGF and QME methods can be exercised in a compatible manner,
to provide nonequilibrium quantities, particularly,
 {\it high order cumulants} of the current
at {\it strong}  metal-molecule coupling.
This is achieved by working out the methods
consistently, namely, by including the same inelastic scattering processes.

\begin{figure}
\hspace{4mm}
\includegraphics[width=7cm]{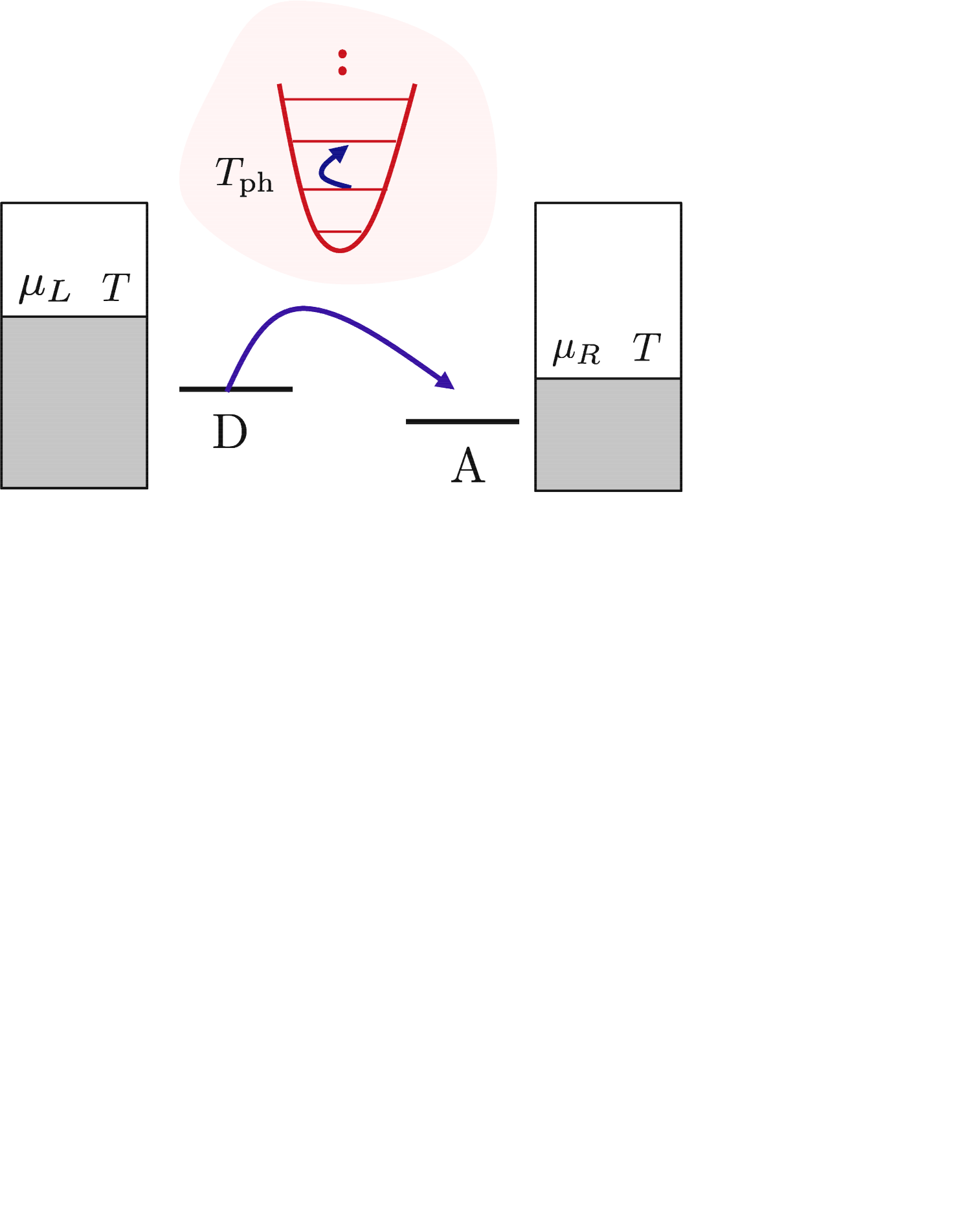}
\caption{Scheme of a biased donor-acceptor molecular junction
with electron hopping between sites
coupled to a (primary) harmonic vibrational mode,
itself embedded in a (secondary) phononic environment.
}
\label{Fig1}
\end{figure}

Our model includes two electronic levels, denoted as
'donor' and 'acceptor', following the chemistry literature on electron transfer.
Charge transfer between these sites takes place via electron interaction with
a local vibrational mode (primary phonon mode), e.g., a torsional mode within a
biphenyl molecule \cite{Markussen}.
The primary mode is further coupled to a secondary phonon bath, allowing it to dissipate its excess energy.
For a schematic representation see Fig. \ref{Fig1}.
The donor-acceptor model was employed by Aviram and Ratner for
proposing a molecular electronic diode \cite{Ratner-Aviram}.
Recently, it was applied for exploring heating effects in molecular junctions
\cite{Lu,SiminePCCP,INFPI3}, and quantum interference effects in molecular conduction \cite{Markussen}
and  thermoelectricity \cite{SimineTE}.

In our treatment, we incorporate the molecule-metal hybridization exactly via a diagonalization procedure
for the electronic part of the Hamiltonian, then perform a perturbation expansion
for the electron-phonon interaction Hamiltonian.
Following the QME approach, steady state observables are readily obtained,
most importantly, with the fluctuation symmetry immediately satisfied \cite{FR}.
In contrast, in the NEGF scheme it is crucial
to employ the so-called random phase approximation (RPA) so as
to take into account corresponding scattering processes and satisfy the fluctuation theorem.

In Ref. \cite{BijayDAPRB,BijayDABeil} we had investigated a similar setup--- without the secondary bath.
However, our focus in \cite{BijayDAPRB,BijayDABeil} has been on comparing transport behavior in two situations:
the case with a harmonic primary model, which was treated with NEGF,
and a model with an anharmonic primary mode (a two-state impurity), handled by QME.
Here, in contrast, 
we focus on reconciling NEGF and the QME techniques by applying them on {\it the same model},
as depicted in Fig. \ref{Fig1}.


The paper in organized as follows.
We introduce our model in Sec. \ref{model}.
In Sec. \ref{meanPh}, we show an agreement between the NEGF and QME methods by calculating the
mean-phonon number.
In Sec. \ref{CGF}, we obtain the CGF for charge and energy current statistics using NEGF,
and explain how cumulants are to be obtained with the QME.
We demonstrate with numerical simulations the equivalence between the two schemes for the first three current cumulants
in Sec. \ref{numerics}. We summarize our work in Sec.\ref{Summary}.

\section{Model}
\label{model}

We consider a minimal two-site molecule with the electronic states identified as
donor (D) and acceptor (A).
The molecules is placed between two metal leads including non-interacting electrons.
Electron transfer between the D and A sites takes place while exchanging energy
with a  molecular vibrational mode, designated as the ``primary phonon". This mode
is coupled to a thermal environment including ``secondary phonons".
We write down the total Hamiltonian as (we set $\hbar = k_B=e=1$ throughout the paper),
\bea
\hat H_T= \hat H_{\rm el} + \hat H_{\rm vib} + \hat H_{\rm  el-vib},
\label{eq:Htotal}
\eea
with
\bea
\hat H_{\rm el}&=& {\epsilon}_d \hat c_d^{\dagger} \hat c_d + {\epsilon}_a \hat c_a^{\dagger} \hat c_a
+ \sum_{l\in L} {\epsilon}_l \hat c_l^{\dagger} \hat c_l
+ \sum_{r\in R} {\epsilon}_r \hat c_r^{\dagger} \hat c_r
\nonumber\\
&+& \sum_{l \in L } v_{l} (\hat c_l^{\dagger} \hat c_d \!+\! \hat c_d^{\dagger} \hat c_l) \!+\! \sum_{r \in R} v_{r} (\hat c_r^{\dagger} \hat c_a \!+\! \hat c_{a}^{\dagger} \hat c_r),
\label{eq:Hel}
\eea
where $\epsilon_{d}, \epsilon_{a}$ are the donor and acceptor site energies, coupled to
the left $L$ and right $R$ metal leads by real-valued hopping elements $v_l$ and $v_r$, respectively.
$\hat c^{\dagger}$ and $\hat c$ are fermionic creation and annihilation operators for the respective regions.
$\hat H_{\rm vib}$ is the Hamiltonian for the vibrational degrees of freedom. It
consists the primary phonon of frequency $\omega_0$, the secondary phonon bath, and a real-valued linear coupling term
with matrix elements $\nu_{j}$,
\be
\hat H_{\rm vib} = \omega_0  \hat b_{0}^{\dagger} \hat b_0 +
\sum_{j} \omega_{j} \hat b_{j}^{\dagger} \hat b_{j} + (\hat b_0^{\dagger} + \hat b_0) \sum_{j} \nu_{j} (\hat b_{j}^{\dagger} + \hat b_{j}).
\label{eq:Hvib}
\ee
The interaction between electrons in the junction and the primary mode is given by the ``off-diagonal" model,
\be
\hat H_{\rm el-vib} = g [\hat c_{d}^{\dagger} \hat c_{a} + \hat c_{a}^{\dagger} \hat c_d ]  (\hat b_0^{\dagger} + \hat b_0).
\label{eq:Hevib}
\ee
Here $\hat b_0(\hat b_0^{\dagger})$ and $\hat b_{j}(\hat b_{j}^{\dagger})$ are bosonic annihilation (creation) operators for
vibrational modes with frequencies $\omega_0$ and $\omega_{j}$, respectively.
The model has been justified in details in Ref. \cite{BijayDAPRB}.
The quadratic electron Hamiltonian (\ref{eq:Hel})
can be diagonalized via a unitary transformation \cite{SiminePCCP}.
The total Hamiltonian now reads
\bea
\hat H_T&=&   \hat H_{\rm vib} + \sum_{l} \epsilon_l \hat a_l^{\dagger} \hat a_l + \sum_{r} \epsilon_r \hat a_r^{\dagger} \hat a_r
\nonumber \\
& + & \sum_{l\in L,r\in R} g \big[\gamma_l \gamma_r^* \hat a_l^{\dagger} \hat a_r \!+\! h.c.\big] (\hat b_0^{\dagger} \!+\! \hat b_0),
\label{eq:tot-h}
\eea
where the new fermionic operators $\hat a_{l,r}$ are related to the original operators according to
\bea
\hat c_d &=& \sum_{l} \gamma_l \hat a_l, \quad \hat c_a = \sum_{r} \gamma_r \hat a_r, \nonumber \\
\hat c_l &=& \sum_{l'} \eta_{l l'} \hat a_{l'}, \quad \hat c_r =\sum_{r'} \eta_{r r'} \hat a_{r'},
\label{eq:newoperators}
\eea
with the dimensionless coefficients
\be
\gamma_l = \frac{v_l}{\epsilon_l-\epsilon_d - \sum_{l'} \frac{v_{l'}^2}{\epsilon_l-\epsilon_{l'} + i \delta}},
\quad \eta_{ll'} = \delta_{ll'}- \frac{v_l \gamma_{l'}}{\epsilon_l-\epsilon_{l'} + i \delta}.
\label{eq:trans}
\ee
Here $\delta$ is a positive infinitesimal number introduced to ensure causality.
Analogous expressions can be written for the $r$ set.
The expectation values for the e.g. $l$ number operators satisfy
$\langle \hat a_{l}^{\dagger} \hat a_{l'} \rangle = \delta_{l l'} f_L(\epsilon_{l})$
with $f_{L}(\epsilon_{l})= \big\{\exp[\beta_{L} (\epsilon_{l}-\mu_{L})] +1\big\}^{-1}$
the Fermi function of the left lead with the chemical potential $\mu_L$ and temperature $T_L=1/\beta_L$.

The primary vibration suffers dissipation due to its coupling to the bosonic-phononic
and fermionic-electronic environments.
We now organize Eq. (\ref{eq:tot-h}) in the following form, $\hat H_T=\hat H_0+\hat V$.
$\hat H_0$ comprises the non-interacting terms of the model (primary and secondary modes, electron baths), and
$\hat V$ includes the interaction Hamiltonian, between the primary mode and the different baths,
\bea
\hat V = (\hat b_0^{\dagger}+\hat b_0) (\hat B_{\rm el} + \hat B_{\rm ph}).
\label{eq:V}
\eea
The baths' operators are,
\bea
\hat B_{\rm el}&=&
g \Big[ \sum_{l,r} \gamma_{l}^{*} \gamma_r \hat a_{l}^{\dagger} \hat a_{r} + \gamma_{l}\gamma_r^{*}  \hat a_{r}^{\dagger} \hat a_{l} \Big],
\nonumber\\
\hat B_{\rm ph} &=& \sum_{\alpha} \nu_{j} (\hat b_{j}^{\dagger} + \hat b_{j}).
\label{eq:Bs}
\eea
%
%
The following expressions are central to our discussion below,
\bea
k_{d}^{\rm el/{\rm ph}}&=& \int_{-\infty}^{\infty} d\tau e^{-i \omega_0 \tau}  \langle \hat B_{\rm el/{\rm ph}}(0)\, \hat B_{\rm el/{\rm ph}}(\tau) \rangle, \nonumber \\
k_{u}^{\rm el/{\rm ph}}&=& \int_{-\infty}^{\infty} d\tau e^{i \omega_0 \tau}  \langle \hat B_{\rm el/{\rm ph}}(0) \, \hat B_{\rm el/{\rm ph}}(\tau)\rangle,
\label{eq:kF}
\eea
defining excitation (u) and relaxation (d) rate constants within the special-primary harmonic mode,
 between neighboring states,
driven by either the electronic or the phononic reservoirs.
In the above definition, the electronic rates are computed by taking an average with respect to the canonical state of the
left and right metal leads, $\hat \rho_{\rm el}=\hat \rho_L\hat \rho_R$,
with  $\hat \rho_{\alpha}=e^{-\beta_{\alpha} (\hat H_{\alpha}-\mu_{\alpha}\hat N_{\alpha})}/Z_{\alpha}$; $\alpha=L,R$.
$Z_{\alpha}$ is the partition function and $\hat H_{L,R}$ are the Hamiltonians for the left and right compartments,
second and third terms in Eq. (\ref{eq:tot-h}).
The phonon-bath induced rates are evaluated with the average taken over the canonical distribution
$\hat \rho_{\rm ph}=e^{-\beta{\rm ph}\hat H_{\rm ph}}/Z_{\rm ph}$ with
$\hat H_{\rm ph}=\sum_{j}\omega_j\hat b_j^{\dagger}\hat b_j$, and the inverse temperature $\beta_{\rm ph}=1/T_{\rm ph}$.
The operators are written in the interaction representation, $ \hat B(\tau)=e^{i\hat H_0\tau}\hat Be^{-i\hat H_0\tau}$.
Using Eq. (\ref{eq:Bs}), it can be shown that
\bea
k^{\rm el}_{u/d}= [k^{\rm el}_{u/d}]^{L \to R} + [k^{\rm el}_{u/d}]^{R \to L},
\eea
where the directional rates are  \cite{SiminePCCP}
\bea
\big[k^{\rm el}_{u}\big]^{L \to R} &=& \int_{-\infty}^{\infty}\! \frac{d\epsilon}{2\pi} f_L(\epsilon) (1\!-\!f_R(\epsilon\!-\!\omega_0)) J_L(\epsilon) J_R(\epsilon\!-\!\omega_0), \nonumber \\
\big[k^{\rm el}_{u}\big]^{R \to L} &=&  \int_{-\infty}^{\infty} \!\frac{d\epsilon}{2\pi} f_R(\epsilon) (1\!-\!f_L(\epsilon\!-\!\omega_0)) J_R(\epsilon) J_L(\epsilon\!-\!\omega_0).
\nonumber\\
\big[k^{\rm el}_{d}\big]^{L \to R} &=& \int_{-\infty}^{\infty}\! \frac{d\epsilon}{2\pi} f_L(\epsilon) (1\!-\!f_R(\epsilon\!+\!\omega_0)) J_L(\epsilon) J_R(\epsilon\!+\!\omega_0), \nonumber \\
\big[k^{\rm el}_{d}\big]^{R \to L} &=&  \int_{-\infty}^{\infty} \!\frac{d\epsilon}{2\pi} f_R(\epsilon) (1\!-\!f_L(\epsilon\!+\!\omega_0)) J_R(\epsilon) J_L(\epsilon\!+\!\omega_0).
\nonumber \\
\label{eq:kel}
\eea
Here $J_{\alpha}(\omega)$ are spectral functions, of Lorentzian form, centered around
the donor (${\epsilon}_d)$ and acceptor (${\epsilon}_a$) site energies with the broadening
$\Gamma_{\alpha}(\epsilon)=2\pi\sum_{k\in\alpha}v_k^2\delta(\epsilon_k-\epsilon)$,
\bea
J_L(\epsilon) &=& g \frac{\Gamma_L(\epsilon)}{(\epsilon-\epsilon_d)^2 + \Gamma_L(\epsilon)^2/4}, \nonumber \\
J_R(\epsilon) &=& g \frac{\Gamma_R(\epsilon)}{(\epsilon-\epsilon_a)^2 + \Gamma_R(\epsilon)^2/4}.
\label{eq:J}
\eea
The electronic rates (\ref{eq:kel}) are nonzero when (i)  both leads are not fully occupied or empty,
 and (ii) the overlap between the spectral functions, differing by one quanta of energy, is non-negligible.
%
The phonon-bath induced rates can be similarly evaluated,
\bea
k_{u}^{\rm ph}=\Gamma_{\rm ph}(\omega_0)n_{\rm ph}(\omega_0), \,\,
k_d^{\rm ph}=\Gamma_{\rm ph}(\omega_0)\left[1+n_{\rm ph}(\omega_0)\right],
\nonumber\\
\label{eq:kph}
\eea
with the coupling energy
\bea
\Gamma_{\rm ph}(\omega)=2\pi\sum_{j}\nu_j^2\delta(\omega-\omega_j),
\eea
and the Bose-Einstein occupation factor
$n_{\rm ph}(\omega)=\left[e^{\beta_{\rm ph}\omega}-1\right]^{-1}$.
We also define the total rates which determine the dynamics of the primary mode,
\bea
k_d=k_d^{\rm el} + k_d^{\rm ph},\,\,\,\
k_u=k_u^{\rm el} + k_u^{\rm ph}.
\label{eq:ktotal}
\eea
Appendix A generalizes these definitions to include counting fields for charge and energy.
In the next sections we calculate several observables far from equilibrium using the NEGF and QME methods, and
demonstrate their agreement when the NEGF method is carefully performed to include all single-phonon electron scattering processes.


\section{Mean phonon number}
\label{meanPh}

We compute here the mean phonon number for the primary mode,
\bea
\langle \hat n \rangle \equiv \langle \hat b_0^{\dagger} \hat b_0 \rangle.
\eea
We employ the NEGF technique under the RPA approximation, then the QME approach under the Born-Markov and
the secular approximations.
Results are valid under the assumption of {\it weak} (second-order) electron-primary phonon interaction $g$,
as well as weak primary-secondary phonon couplings $\nu_j$.

\begin{figure} [pb]
\includegraphics[width=8cm]{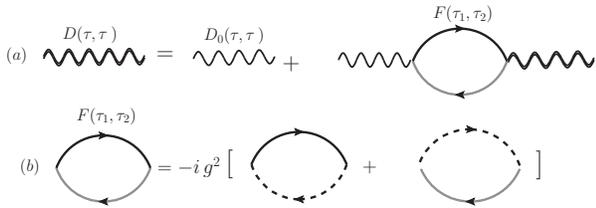}
\caption{(a) Dyson equation for the phonon Green's function
$D(\tau,\tau')$ in contour time.
(b) Electron-hole propagator $F(\tau,\tau')$.
The solid and dashed lines represent the unperturbed Green's functions $g_l$ and $g_r$ for the leads, respectively.}
\label{Feynman_phonon}
\end{figure}

\subsection{NEGF approach}

We define the contour-ordered phonon Green's function as
\bea
D(\tau,\tau') &\equiv& -i \langle T_c \hat X(\tau) \hat X(\tau') \rangle, \nonumber \\
&=& -i \langle T_c \hat X_I(\tau) \hat X_I(\tau') e^{-i/\hbar \int d\tau_1 \hat H^I_{\rm el-vib}(\tau_1)}\rangle
\eea
where $\hat X= (\hat b_0+\hat b_0^{\dagger})$ is proportional to the primary phonon displacement operator.
$T_c$ is the contour-ordered operator responsible for rearrangement of the operators according to their contour time.
In the second line, the operators are written in the interaction picture 
with respect to the noninteracting parts $\hat H_{\rm el}+\hat H_{\rm vib}$, the quadratic parts of the total Hamiltonian.
The perturbative expansion generates terms of different orders in the electron-phonon coupling $g$,
(but it is exact to all orders so far in the coupling of the primary phonon to the phonon bath).
A naive perturbative calculation with diagrams up to a particular order of $g$ leads to the
{\it violation} of different symmetries preserving physical processes,
such as the conservation of charge and energy currents.  
In order to restore basic symmetries, one has to sum over an infinite-subclass of diagrams, taking into account
all electron scattering processes  
which are facilitated by the absorption or emission of a single quanta $\omega_0$.
This can be done by employing the so-called random phase approximation (RPA) \cite{Utsumi2, Atland}
where a particular type of ring diagrams are summed over,  see Fig.~(\ref{Feynman_phonon}).
We can represent this infinite summation in a closed Dyson-like (kinetic) equation for $D(\tau,\tau')$,
\bea
D(\tau,\tau')&=& D_0(\tau,\tau')  \nonumber \\
&\!+\!& \int d\tau_1 \int d\tau_2 D_0(\tau,\tau_1) F(\tau_1,\tau_2) D(\tau_2,\tau').
\eea
$D_0$ is the primary phonon's Green's function, and it includes the effect of the secondary phonon bath.
$F$, the electron-hole propagator, involves both left and right lead electron Green's function (unperturbed), and it
describes electron hopping processes from $L$ to $R$ and vice versa, assisted by scatterings with phonon modes.
It is given as
\bea
F(\tau_1,\tau_2) &=& -i g^2 \,\sum_{l\in L,r \in R} |\gamma_l|^2 \gamma_r|^2 \big[ g_l (\tau_1,\tau_2) \, {g}_r(\tau_2,\tau_1)  \nonumber \\
&+& {g}_r(\tau_1,\tau_2) \, g_l(\tau_2,\tau_1)\big].
\label{eh-prop}
\eea
This function is symmetric under the exchange of the contour time parameters $\tau_1$ and $\tau_2$.
The free (unperturbed) electronic Green's functions are
\bea
g_{l}(\tau_1,\tau_2)\!&=&\! - i \,\langle T_c \hat a_{l}(\tau_1) \hat a_{l}^{\dagger}(\tau_2)\rangle_L, \nonumber \\
{g}_{r}(\tau_1,\tau_2)\!&=&\! - i \langle T_c {\hat a}_{r}(\tau_1) {\hat a}_{r}^{\dagger}(\tau_2)\rangle_R,
\label{free}
\eea
with the average performed over the respective grand canonical distribution functions
with well-defined temperatures $T_{\alpha}$ and chemical potentials $\mu_{\alpha}$, $\alpha=L,R$.
In the long-time (steady state) limit, different real-time components of $D$ can be obtained.
The convolution in time domain results in a  multiplicative form in frequency domain,
\begin{small}
\bea
&&{\bf D}(\omega) =\begin{bmatrix}
                         D^{t}(\omega)& {D}^{<}(\omega)   \\
                          D^{>}(\omega) & D^{\bar{t}}(\omega)
\end{bmatrix} =\big[{\bf D}_0^{-1}(\omega)- {\bf F}(\omega)\big]^{-1} =
\nonumber \\
\!\!\!\!&&
\begin{bmatrix}
                         [D_0^r]^{-1}(\omega)\!-\!\Sigma_{\rm ph}^{<}(\omega)\!-\!F^t(\omega) \!& \!\!\Sigma_{\rm ph}^{<}(\omega)\!+\!{F}^{<}(\omega)   \\
                          \Sigma_{\rm ph}^{>}(\omega)\!+\!{F}^{>}(\omega) \!&\!\! \!-\![D_0^r]^{-1}(\omega)\!-\!\Sigma_{\rm ph}^{>}(\omega)\!-\!F^{\bar{t}}(\omega)
\end{bmatrix}^{-1}, \nonumber \\
\eea
\end{small}
where $t, \bar{t}, <, >$ are the time ordered, anti-time ordered, lesser and greater components of the Green's function.
The primary phonon retarded Green's function  is
\bea
[D_0^r(\omega)]^{-1}=(\omega^2-\omega_0^2)/2\omega_0-\Sigma_{\rm ph}^r
\eea
with $\Sigma_{\rm ph}$ the self energy due to the coupling of the mode to the phonon bath.
For the mean phonon number, we are interested in the lesser $D^{<}(\omega)$ and greater $D^{>}(\omega)$ components.
This can be readily calculated by inverting the $2\times 2$ matrix
which results in
\bea
D^{</>}(\omega)&=& \frac{\Pi^{</>}(\omega)}{\Big[\frac{\omega^2-\omega_0^2}{2 \omega_0} -\Big(\frac{\Pi^t\!-\!\Pi^{\bar{t}}}{2}\Big)\Big]^2+ A_{0}(\omega)},
\label{less-greater-eq}
\eea
with  the total self-energy  due to both electronic and phononic baths,
\bea
\Pi^{</>}(\omega)&=& \Sigma_{\rm ph}^{</>}(\omega)+ F^{</>}(\omega),  \nonumber\\
\Pi^{t/\bar t}(\omega)&=& \Sigma_{\rm ph}^{t/\bar t}(\omega)+ F^{t/\bar t}(\omega).
\eea
The function $A_0(\omega)$ is written solely in terms of $\Pi^{</>}(\omega)$ as
\be
A_{0}(\omega)= -\frac{1}{4} \left[\Pi^>(\omega) \!-\! \Pi^{<}(\omega)\right]^2,
\label{central-quantity}
\ee
and it will emerge as the central quantity in this problem.
The mean-square displacement, in steady-state, can be readily
obtained from the phonon Green's function as
\bea
\langle \hat X^2 \rangle &=& \frac{i}{2} \int_{-\infty}^{\infty} \frac{d\omega}{2 \pi} \Big[D^{<}(\omega)\!+\!D^{>}(\omega)\Big], \nonumber \\
&=& \frac{i}{2} \int_{-\infty}^{\infty} \frac{d\omega}{2 \pi} \frac{\Pi^{<}(\omega)+ \Pi^{>}(\omega)}{\Big[\frac{\omega^2-\omega_0^2}{2 \omega_0} -\Big(\frac{\Pi^t\!-\!\Pi^{\bar{t}}}{2}\Big)\Big]^2+ A_{0}(\omega)}.
\label{eq:X2}
\eea
The integration can be performed to include terms
to the lowest nontrivial order in the electron-phonon coupling (order $g^2$) and in $\Gamma_{ph}$.
We employ the residue theorem to perform the integration.
The poles are located at (correct up-to the second order)
$\pm \{\omega_0 + {\rm Re} [\Pi^R(\omega_0)] \pm i \sqrt{A_0(\omega_0)}\}$. The integration in (\ref{eq:X2}) then results in
\be
\int_{-\infty}^{\infty} \frac{d\omega}{2\pi} D^{</>}(\omega) = \frac{\Pi^{</>}(\omega_0)}{\sqrt{A_0(\omega_0)}}.
\ee
Using the expressions for the free (unperturbed) left and right lead Green's functions
in Eq. (\ref{free}), we identify the elements $F^{</>}, \Sigma_{\rm ph}^{</>}$
in terms of mode excitation and relaxation rates as induced by the electronic and phononic baths,
(\ref{eq:kel}) and (\ref{eq:kph}), respectively.
Explicitly, at the vibrational frequency $\omega_0$, we get \cite{BijayDAPRB}
\bea
F^>(\omega_0)= -i k^{\rm el}_{d},\,\,\,\,\,
F^<(\omega_0)= -i k^{\rm el}_{u}.
\eea
and
\bea
\Sigma_{\rm ph}^{>}(\omega_0) =  - i k^{\rm ph}_{d},\,\,\,\,\,
\Sigma_{\rm ph}^{<}(\omega_0)= - i k^{\rm ph}_{u},
\label{phonon-rate}
\eea
%
From Eq.~(\ref{central-quantity}) we obtain
$\sqrt{A_{0}(\omega_0)}= \frac{1}{2} (k_d-k_u)$ assuming $k_d > k_u$,
with the total rates defined in Eq. (\ref{eq:ktotal}).
The final result is thus
\be
\langle \hat X^2 \rangle =  \frac{ k_d + k_u} {k_d - k_u},
\ee
and the mean phonon number is given by
\be
\langle \hat n \rangle = \frac{1}{2} \big[\langle \hat X^2 \rangle -1 \big] = \frac{k_u} {k_d -k_u}.
\label{eq:nNEGF}
\ee

\subsection{QME approach}

We compute here the mean phonon number for the primary mode following a QME- projection operator technique.
First, we identify the baths---to be traced over.
Singling out the primary phonon as the system of interest,
our model system (Fig. \ref{Fig1}) includes two environments:
the electronic degrees of freedom prepare a nonequilibrium electronic bath,  and
the secondary phonons constitute a second thermal bath.

In standard QME approaches, the perturbation expansion is done with respect to the metal-molecule coupling \cite{Mitra}.
Here, instead, we treat the interaction of the primary mode with the electron and phonon baths as weak-perturbative,
but the hybridization of the D and A electronic states to the metals is
treated exactly, with the help of the diagonalization procedure.
In second order of perturbation theory with respect to $g$ and $\nu_j$, and after a Markov approximation,
the reduced density matrix equation for the primary harmonic
mode of frequency $\omega_0$ takes the following form \cite{BookOQS}
\bea
&&\dot{\rho}_s(t)= - i \big[\hat V(t), \rho_s(0)\big] \nonumber \\
&&- \int_{0}^{\infty} d\tau {\rm Tr}_{\rm el, \rm ph} \big\{\big[\hat V(t),\big[\hat V(\tau),
\rho_s(t) \otimes \hat \rho_L \hat \rho_R \otimes \hat \rho_{\rm ph}\big]\big]\big\},
\nonumber\\
\eea
with the operators written in the interaction representation with respect to $\hat H_0$.
The trace is performed over the electronic
and the secondary-phonon bath degrees of freedom, with initial conditions given by
 equilibrium density matrices, as explained below Eq. (\ref{eq:kF}).

Further imposing the secular approximation, i.e., decoupling the evolution of diagonal and off-diagonal elements, 
we obtain a kinetic-type equation for the populations of the primary mode, ${p}_n\equiv \langle n|\rho_s(t)|n\rangle$,
\be
\dot{p}_{n}= -\big[n\, k_{d}+(n+1) \,k_{u}\big] \,{p}_{n} + (n+1) \,k_{d} \,{p}_{n+1} + n \, k_{u} \,{p}_{n-1},
\ee
with the rates (\ref{eq:ktotal}).
The steady state solution, $\dot{p}_n=0$,
is obtained by using a trial form $p_n = c\, y^n $,
which results in the normalized solution \cite{Segal-QME}
\be
p_n = \Big(1- \frac{k_u}{k_d}\Big) \Big(\frac{k_u}{k_d}\Big)^n.
\label{ss-sol}
\ee
The average  phonon number can now be directly evaluated,
\be
\langle \hat n \rangle = \sum_{n} n p_n= \frac{k_u} {k_d -k_u}.
\label{eq:nQME}
\ee
This result precisely matches the NEGF expression (\ref{eq:nNEGF}).
The QME derivation reveals that this result does not rely on
quantum coherence effects within the primary mode, which
may contribute only in higher orders of the electron-phonon coupling energy.

\section{Charge current statistics}
\label{CGF}

We study the statistics (distribution) of charge and energy currents from
the so-called characteristic function,
which is related to the distribution function via a Fourier transform.
Following the two-time measurement procedure \cite{Huanan}, we define the characteristic
function as \cite{FR, Hanggi-review, bijay-ballistic},
\be
{\cal Z}(\lambda_e, \lambda_p) = \Big\langle e^{i \lambda_{e} \hat H_R + i \lambda_{p} \hat N_R} \, e^{-i \lambda_{e} \hat H_R^H(t) - i \lambda_{p} \hat N_R^H(t)} \Big\rangle.
\label{eq:Z}
\ee
Here, $\lambda_e$ and $\lambda_p$ are counting fields for energy and particles, respectively.
$\langle \cdots \rangle$ represents an average with respect to the total density matrix
at the initial time, $\hat \rho_T(0)= \hat \rho_L(0) \hat \rho_R(0)\otimes \hat \rho_{\rm ph}(0)\otimes \rho_{s}(0)$,
a factorized-product form for the three baths (two metals, secondary phonons) and the primary mode (system, denoted by $s$).
The baths are prepared in thermal equilibrium, as explained below Eq. (\ref{eq:kF}).
The operators are written here in the Heisenberg representation
and $\hat H_R=\sum_{r}\epsilon_r \hat a_r^{\dagger}\hat a_r$, $\hat N_R=\sum_r \hat a_r^{\dagger}\hat a_r$.
The CGF is defined as follows,
\bea
{\cal G}(\lambda_e,\lambda_p)&=&  
\lim_{t \to \infty} \frac{1}{t}
\ln {\cal Z}(\lambda_e,\lambda_p).
\label{eq:CGF}
\eea
The steady state current and higher order cumulants can be readily obtained
by taking derivatives with respect to the counting fields.
Specifically, the cumulants of the particle current are given as
\bea
C_m&\equiv&  \langle I_p^{m}\rangle 
\nonumber\\
&\equiv& \frac{\partial^m {\cal G}(\lambda_e,\lambda_p)}{\partial (i \lambda_p)^m}\Big{|}_{\lambda=0}, \,\,\,\, m=1,2,...
\eea
where $\lambda=(\lambda_e,\lambda_p)$. In Sec. \ref{CGF-NEGF} we provide an analytic expression for the CGF (\ref{eq:CGF}) from an NEGF formalism. This allows us to
reach closed-form expressions for the current cumulants,
given in terms of counting-field dependent rates.
Using the QME, in Sec. \ref{CGF-QME} we
directly-numerically compute cumulants, by using a recursive algorithm.
Numerical results as presented in Sec. \ref{numerics} confirm that the two methods yield equivalent results for the first three cumulants.

\subsection{NEGF approach}
\label{CGF-NEGF}

We had recently derived an analytic expression for the CGF for
the model of Fig. \ref{Fig1}---without a secondary phonon bath---using an NEGF approach, by following an RPA scheme  \cite{BijayDAPRB}
We now describe how to generalize this study to include this additional dissipation channel.
It can be shown that  \cite{NEGF-review}
\bea
\ln {\cal Z}_{\rm RPA}(\lambda_e,\lambda_p)
= -\frac{1}{2}\, {\rm Tr}_{\tau} \,\ln \big[I\!-\! D_0(\tau,\tau') \tilde{F}(\tau,\tau')\big],
\nonumber\\
\label{eq:CGF-RPA}
\eea
where tilde refers to $\lambda=(\lambda_e,\lambda_p)$-dependent quantities.
$D_0$ follows the same definition as before, but
the electron-hole propagator is now modified due to the counting fields  \cite{BijayDAPRB},
\bea
\tilde{F}(\tau_1,\tau_2) &=& -i g^2 \,\sum_{l\in L,r \in R} |\gamma_l|^2 \gamma_r|^2 \big[ g_l (\tau_1,\tau_2) \tilde{g}_r(\tau_2,\tau_1)  \nonumber \\
&+& \tilde{g}_r(\tau_1,\tau_2) g_l(\tau_2,\tau_1)\big].
\label{modified-eh-prop}
\eea
%
\begin{figure}
\includegraphics[width=8cm]{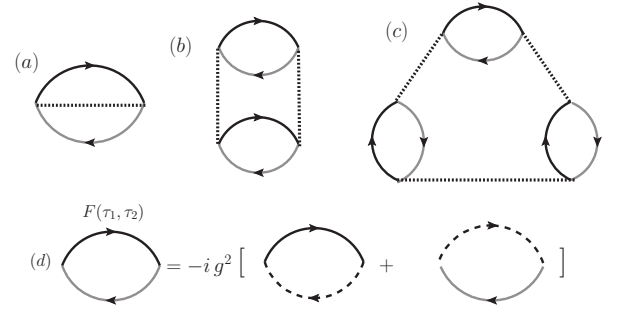}
\caption{Ring-type Feynman diagrams in contour time.
(a) Second-order, (b) fourth-order, and (c) sixth-order diagrams in the electron-phonon coupling.
The dotted line represents the phonon Green's function $D_0$. 
(d) The closed loops are the electron-hole propagator $\tilde{F}$ with solid and dotted lines corresponding
to bare left and right-lead Green's functions.}
\label{Feynman_diagrams}
\end{figure}
The greater and lesser components of $\tilde{F}$ can be expressed by the counting-fields dependent rates,
\bea
\tilde{F}^{>/<}(\omega_0)= - i \tilde{k}^{\rm el}_{d/u},
\eea
see definition in the Appendix. Repeating the procedure of Ref. \cite{BijayDAPRB}---with an additional-secondary phonon bath---we receive the CGF
\be
{\cal G}(\lambda) = \frac{1}{2}(k_{d} - k_{u}) - \frac{1}{2}\sqrt{(k_{u} + k_{d})^2 - 4 \,\tilde{k}_{d}\, \tilde{k}_{u} }.
\label{CGF-H}
\ee
This expression is correct up to second-order in the primary phonon-electron and  primary phonon-secondary phonon couplings.
The charge current immediately follows,
\be
\langle I_p \rangle \equiv \frac{\partial {\cal G}(\lambda)}{\partial (i \lambda_p)}\Big{|}_{\lambda=0} =\, \frac{\big(k^{\rm el}_{u}\big)'k_d + \big(k^{\rm el}_{d}\big)' k_u}{k_{d} \!-\! k_{u}},
\label{eq:current-NEGF}
\ee
with $\big(k^{\rm el}_{u/d}\big)'$ the difference between electronic rates, see Eq. (\ref{eq:klamdiff}).
The particle current can also be expressed in terms of mean phonon number, or the mean-square displacement, as
\bea
\langle I_p \rangle  &=&\, \big(k^{\rm el}_{u}\big)' (1+\langle \hat n \rangle) + \big(k^{\rm el}_{d}\big)' \langle \hat n \rangle,
\nonumber\\
&\!=\!&\langle \hat X^2 \rangle \Big[\big(k^{\rm el}_{u}\big)' \!+\! \big(k^{\rm el}_{d}\big)'\Big] \!+\! \frac{1}{2}\,\Big[\big(k^{\rm el}_{u}\big)'\! -\! \big(k^{\rm el}_{d}\big)'\Big].
\label{eq:avg-current-phonon}
\eea
The fluctuation (noise) in the charge current can be similarly organized,
\bea
\langle S_p \rangle \equiv  \frac{\partial^2 {\cal G}(\lambda)}{\partial (i \lambda_p)^2}\Big{|}_{\lambda=0}
&=& \frac{ k_{u}^{\rm el} \,k_{d} + k_{d}^{\rm el} \,k_{u} + 2 (k_u^{\rm el})' (k_d^{\rm el})'}{k_{d} \!-\!  \, k_{u}}
\nonumber \\
& + & \frac{2}{(k_{d} \!-\! \, k_{u})} \langle I_p \rangle^2.
\eea
The first term in $\langle S_p\rangle$ is refereed to as the equilibrium noise (although it depends on the non-equilibrium condition),
the second term adds a purely non-equilibrium contribution.

In the absence of the secondary phonon bath, the above expressions can be simplified to give
\bea
\langle I_p \rangle &=& 2 \, \frac{k_{d} ^{R \to L} k_{u}^{R \to L} - k_{d}^{L \to R} k_{u}^{L \to R}}{k_{d}^{\rm el} - k_{u}^{\rm el}}.
\label{eq:Ip}
\eea
The Fano factor $F\equiv \langle S_p\rangle/\langle I_p\rangle$ now obeys
\bea
F &=& \frac{2\langle I_p \rangle }{k_d^{\rm el}-k_u^{\rm el}}
\nonumber\\
&+& 2\frac {k_d^{R \to L} k_u^{R \to L} + k_d^{L \to R} k_u^{L \to R} } {k_d^{R \to L} k_u^{R \to L} - k_d^{L \to R} k_u^{L \to R}}.
\label{eq:F}
\eea
Note that in the absence of the secondary bath the system may develop the so called
``vibrational instability" effect, an uncontrolled heating of the junction \cite{Lu,SiminePCCP}.
At large bias, we reach $k_d^{\rm el} \sim k_u^{\rm el}$ (and even $k_d^{\rm el} < k_u^{\rm el}$ ),
resulting in the blow-up of current.
This physical instability can be relaxed once excess heat within the primary mode
is allowed to dissipate into secondary phonon modes.
In this case, the denominator in Eq. (\ref{eq:current-NEGF}) is given by
$k_d-k_u= k_d^{\rm el}-k_u^{\rm el}+ \Gamma_{\rm ph}$, which is positive even at high bias, as long as $\Gamma_{\rm ph}$ is large enough.

\subsection{QME approach}
\label{CGF-QME}

The NEGF formalism furnishes a closed-form expression for the CGF.
In the language of kinetic rate equations, this could be achieved by following an approach detailed in Ref.
\cite{vanKampen}. 
Here, in contrast, we calculate  high order cumulants directly from the Liouvillian. This approach is specifically useful as
it is not limited to harmonic systems, and could be implemented to describe junctions with anharmonic modes.

In the QME approach, the FCS can be obtained by following a counting field dependent master equation
for $\rho^{s}_{\lambda}$, the reduced density matrix of the primary mode. It is defined from Eq. (\ref{eq:Z}),
\bea
{\cal Z}(\lambda_e, \lambda_p)
= {\rm Tr_T} \big[\rho^{T}_{\lambda_e,\lambda_p}(t)\big]=
{\rm Tr}_{s} \big[\rho^{s}_{\lambda_e,\lambda_p}(t)\big].
\eea
%
To include charge and energy measurements,
the electron-phonon coupling term in the Hamiltonian (\ref{eq:tot-h})
is dressed by counting fields,
\bea
{\hat V}_{\pm \lambda/2}= (\hat b_0^{\dagger}+\hat b_0) ({\hat B}^{\rm el}_{\pm \lambda/2} \!+\! \hat B_{\rm ph}),
\eea
with
\bea
{\hat B}^{\rm el}_{\pm \lambda/2}=g \Big[ \sum_{l,r} \gamma_{l}^{*} \gamma_r \hat a_{l}^{\dagger} \hat a_{r} e^{ \mp \frac{i}{2} (\lambda_p + \epsilon_r \lambda_e)} + h.c. \Big].
\eea
The counting-field dependent density matrix equation for the primary mode can be written formally as \cite{BijayDAPRB}
%
%
\begin{widetext}
\bea
\dot{{\rho}}^{{\rm s}}_{\lambda}(t)
= -\int_{0}^{t} dt' &&{\rm Tr}_{\rm el,ph}
\Big[{\hat V}_{-\lambda/2}(t) {\hat V}_{-\lambda/2}(t') {\hat \rho}_{\lambda}^T(t') + {\hat \rho}_{\lambda}^T(t') {\hat V}_{\lambda/2}(t') \hat V_{\lambda/2}(t) \nonumber \\
&& - {\hat V}_{-\lambda/2}(t') \hat \rho_{\lambda}^T(t') {\hat V}_{\lambda/2}(t) - \hat {V}_{-\lambda/2}(t) {\hat \rho}_{\lambda}^T(t') {\hat V}_{\lambda/2}(t')\Big].
\eea
\end{widetext}
%
Recall that $\lambda$ is a shorthand notation for both $\lambda_e$ and $\lambda_p$.
As demonstrated in Refs. \cite{SiminePCCP,BijayDAPRB}, under the Born-Markov approximation
and the secular approximation,
we can write down the population dynamics for the vibrational state as
\be
\dot{{p}}^{\lambda}_{n}(t)= -\big[n k_{d}\!+\!(n\!+\!1) k_{u}\big]\,{p}^{\lambda}_{n} \!+\! (n\!+\!1) \, \tilde{k}_{d} \, {p}^{\lambda}_{n\!+\!1} \!+\! n \,\tilde{k}_{u} \, {p}^{\lambda}_{n-1},
\ee
with $p_n^{\lambda}(t)\equiv \langle n|\rho^s_{\lambda}(t)|n\rangle$,
$n=0,1,2,\cdots$.
It is convenient to condense these equations into a matrix form,
\bea
 |\dot p^{\lambda}\rangle = \tilde{\cal L} \, |{p}^{\lambda}\rangle,
\eea
with $\tilde{\cal L}$
the $\lambda$-dependent Liouvillian. The long-time (steady state) limit provides the CGF,
\be
{\cal G}(\lambda) = \lim_{t \to \infty} \frac{1}{t} \ln {\cal Z} (\lambda)
= \lim_{t \to \infty} \frac{1}{t} \ln \langle I|{  {p^{\lambda}}}(t)\rangle,
\label{long-CGF}
\ee
where $\langle I|=(1,1,1,\cdots)^T$ is the identity vector.
It is obvious from the above equation that the steady state CGF is given by the smallest eigenvalue of
the dressed Liouvillian $\tilde{\cal L}$.
However, because of its infinite dimensionality a closed-form expression for the eigenvalue is not feasible to obtain \cite{vanKampen}.
The cumulants can however be calculated numerically following
the Rayleigh-Schr\"dinger perturbation scheme  as described in \cite{Flindt1, Flindt2}.

An analytical expression for the current can be recovered from Eq. (\ref{long-CGF}) using
the steady state solution of the vibrational mode, Eq. (\ref{ss-sol}),
\bea
\langle I_p \rangle &=& \frac{\partial {\cal G}(\lambda)}{\partial (i \lambda_p)}\Big|_{\lambda=0} = \langle I | {\cal L}^{'}(0) |
p_{ss} \rangle, \nonumber \\
&=& (k_d^{\rm el})' (p_1 + 2 p_2 + 3 p_3 + \cdots)
\nonumber\\
&+& (k_u^{\rm el})'(p_0 + 2 p_1 + 3 p_2 + \cdots),
\nonumber \\
&=&
\frac{ \big(k^{\rm el}_{d}\big)' k_u +  \big(k^{\rm el}_{u}\big)'k_d  }{k_{d} \!-\! k_{u}},
\label{current-QME}
\eea
where
%
\bea
&&{\cal L}^{'}(0)\equiv \frac{\partial \tilde{\cal L}}{\partial {(i\lambda_p)}}{\Big|}_{\lambda=0}
=
\nonumber\\
&&\begin{bmatrix}
0 & ({k}_d^{\rm el})' & 0 & 0 & ... & ... &... \\
({k}_u^{\rm el})'& 0 & 2({k}_d^{\rm el})' & 0 & 0 & ... & ... \\
0 & 2 ({k}_u^{\rm el})'& 0 & 3({k}_d^{\rm el})' & 0 & ... &... \\
0 & 0 & 3 ({k}_u^{\rm el})'& 0 & 4({k}_d^{\rm el})' & 0 & ... \\
0 & 0 &  0 & ... & ... & ... & ... \\
... & ...&  ... & ... & ... & ... & ...
\end{bmatrix}
\eea
and $|p_{ss}\rangle=|p_0,p_1,p_2, \cdots \rangle$ is the column vector with the steady state populations.
The expression for current, Eq. (\ref{current-QME}), matches with the NEGF result, Eq. (\ref{eq:current-NEGF}).

To calculate higher order statistics, the main inputs are
(i) higher order derivatives of the Liouvillian with respect to counting fields,
(ii) the steady state solution of the vibrational states,
and, (iii) the pseudoinverse matrix, which is explained below.
For example, the second cumulant, or the noise due to charge-fluctuation, is given by \cite{Flindt1,Flindt2}
\bea
\langle S_p \rangle &\equiv&
 \frac{\partial^2 {\cal G}(\lambda)}{\partial (i \lambda_p)^2}\Big|_{\lambda=0}
\nonumber\\
&=&
\langle I|{\cal L}^{''}(0)|p _{ss}\rangle -2 \langle I| {\cal L}^{'}(0) {\cal R} {\cal L}^{'}(0)| p_{ss}  \rangle.
\eea
Here ${\cal R}$ is the pseudoinverse, defined as
${\cal R}= {\cal Q} {\cal L}^{-1} {\cal Q}$ with ${\cal Q}= {\cal Q}^2= 1-{\cal P}$, ${\cal P}={\cal P}^2= |p_{ss}\rangle \langle I|$.
Note that the pseudoinverse is a well-defined quantity; the inverse is performed in the subspace corresponding
to ${\cal Q}$ which excludes the zero eigenvalue of the Liouvillian.
We calculate ${\cal R}$ numerically by computing the eigenvalues and the corresponding left and right eigenvectors of ${\cal L}$,
\be
{\cal R} =  {\cal Q} {\cal L}^{-1} {\cal Q} = \sum_{n \neq 0} \frac{1}{\gamma_n} {\cal Q} |R_{n}\rangle \rangle \langle \langle L_n| {\cal Q},
\ee
where ${\cal L} |R_{n}\rangle  = \gamma_n |R_{n}\rangle , \langle L_{n}| {\cal L} = \gamma_n \langle L_{n}|$.
We have also used the fact that ${\cal Q}|p_{ss}\rangle =0$. Note that the sum excludes the zero eigenvalue.
Similarly, the third cumulant (skewness) can be obtained as \cite{Flindt2}
\bea
 \langle I_p^3\rangle  \!&=&\! \langle I | {\cal L}^{'''}(0) \!+\! 6 {\cal L}^{'}(0) {\cal R} {\cal L}^{'}(0) {\cal R} {\cal L}^{'}(0) \!-\!3 {\cal L}^{''}(0) {\cal R} {\cal L}^{'}(0)\nonumber \\
\!&-&\! 3 {\cal L}^{'}(0) {\cal R} {\cal L}^{''}(0) \!-\! 6 {\cal L}^{'}(0) {\cal R}^2 {\cal L}^{'}(0) \langle I_p \rangle | p_{ss}\rangle.
\eea
%

\section{Results}
\label{numerics}

\subsection{Simulations}

In this section, we present numerical results for the mode occupation, current, and cumulants far from equilibrium.
In presenting these results, we have two objectives in mind:
(i) Demonstrate that results for the cumulants, reached from NEGF and QME, perfectly agree.
(ii) Examine the functionality of the junction under bias and find out what information does
the current and its cumulants convey on the junction's microscopic parameters.

Simulations are performed in the wide-band limit for the the leads. For simplicity, we assume symmetric coupling,
$\Gamma_{L,R}(\epsilon)=\Gamma$.
We set the equilibrium Fermi level at zero and move the bias symmetrically about the Fermi energy
with $\Delta \mu\equiv \mu_R-\mu_L$.
We work with the following representative parameters:
 ${\epsilon}_d\!=\!{\epsilon}_a\!=\!{\epsilon}_0=0.2$ eV as  the energies of the molecular electronic states,
$\Gamma=0.01-0.1$ eV for
the metal-molecule hybridization, $g=0.01$ eV as the electron-primary phonon coupling.
The electronic temperature is set at $T_L=T_R=100$ K, while the phonon bath is set at $T_{\rm ph}=300$ K.
Our conclusions do not depend on this precise choice of parameters.

\begin{figure}
\includegraphics[width=0.5\textwidth]{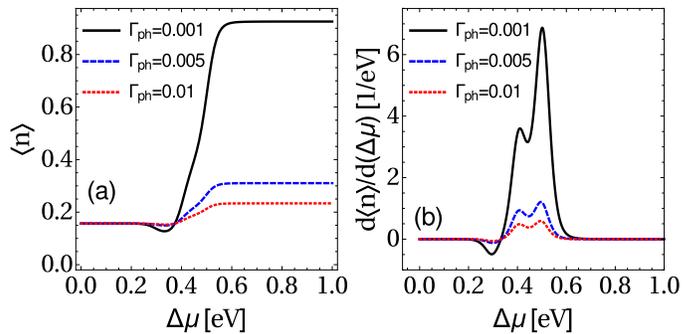}
\centering
\caption{(Color online)
Heating and cooling of the primary mode.
(a) Mean phonon number as a function of bias, and (b) its derivative,
using different phonon damping rates.
Parameters are $\epsilon_d=\epsilon_a=0.2$ eV,
$\omega_0=0.05$ eV, $g=0.01$ eV, $\Gamma=0.01$ eV, $T_L=T_R=100$ K, $T_{\rm ph}=300$ K.}
\label{mean-ph}
\end{figure}

In Fig. \ref{mean-ph}a we study the average phonon number in the primary mode as a
function of the bias difference using $\Gamma < \omega_0$.
The average phonon number increases once the bias $\Delta \mu$  exceeds the value
$2 \epsilon_0 = 0.4$  eV,  revealed by the peak structure in  Fig. \ref{mean-ph}b.
The second peak arises when the bias exceeds the value $2 (\epsilon_0 + \omega_0)$.
With increasing bias, the bath-induced  excitation rate ($k_u$) increases,
resulting in higher values for the average phonon number.
The energy gap between the two peaks matches the value $2\omega_0$.
Notice that the magnitude of the second peak at $\Delta \mu =2 (\epsilon_0 + \omega_0)$
is larger than the first one,
because of the availability of many additional transport channels.
With increasing $\Gamma_{ph}$, the mode dissipates its excess energy to the phonon bath,
resulting in low values for the average phonon number.
Interestingly, when the  phonon damping rate is small, $\Gamma_{\rm ph}=0.001$ eV, a cooling effect takes place
around $\Delta \mu=0.3$ eV.

\begin{figure}
\includegraphics[width=0.5\textwidth]{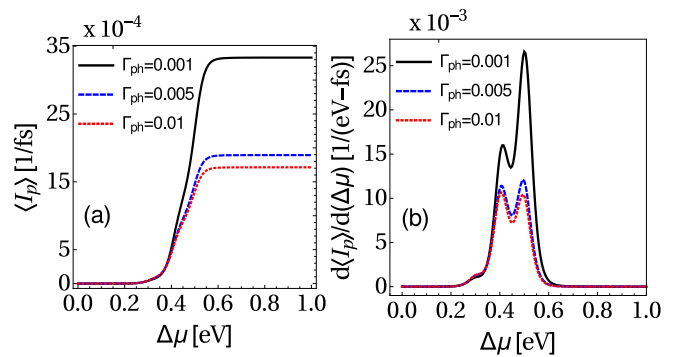}
\centering
\caption{(Color online)
I-V characteristics of the junction.
(a) Current and (b) differential conductance.
Parameters are the same as in Fig. \ref{mean-ph}.
}
\label{current}
\end{figure}

\begin{figure}
\centering
\includegraphics[width=0.45\textwidth]{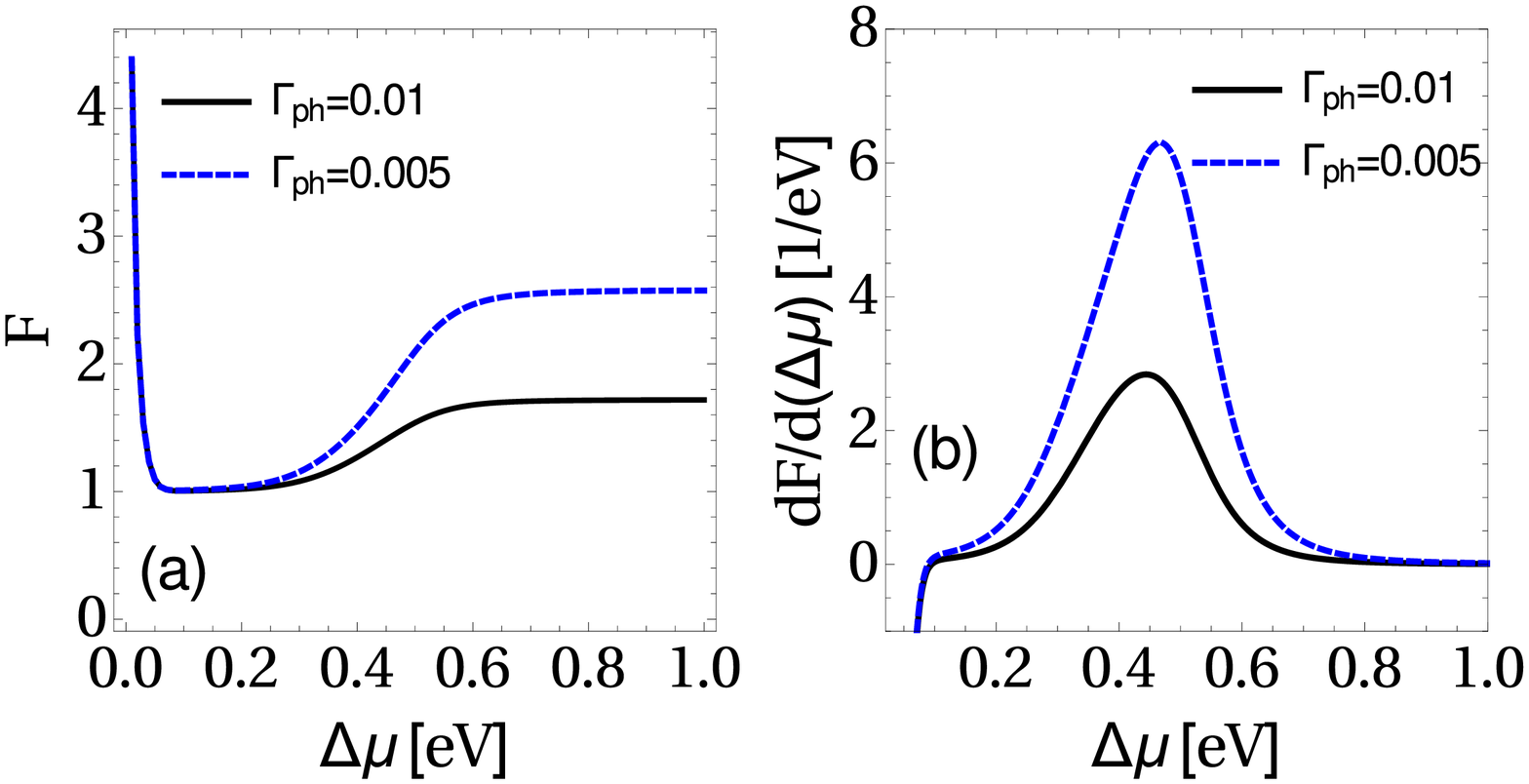}
\caption{(Color online)
Second cumulant for charge transfer as obtained from the QME and NEGF methods (overlapping).
The Fano factor is displayed as a function of bias voltage for $\Gamma=0.1$ eV,
other parameters are same as in Fig. \ref{mean-ph}.
}
\label{fano}
\end{figure}

\begin{figure}
\centering
\includegraphics[width=0.45\textwidth]{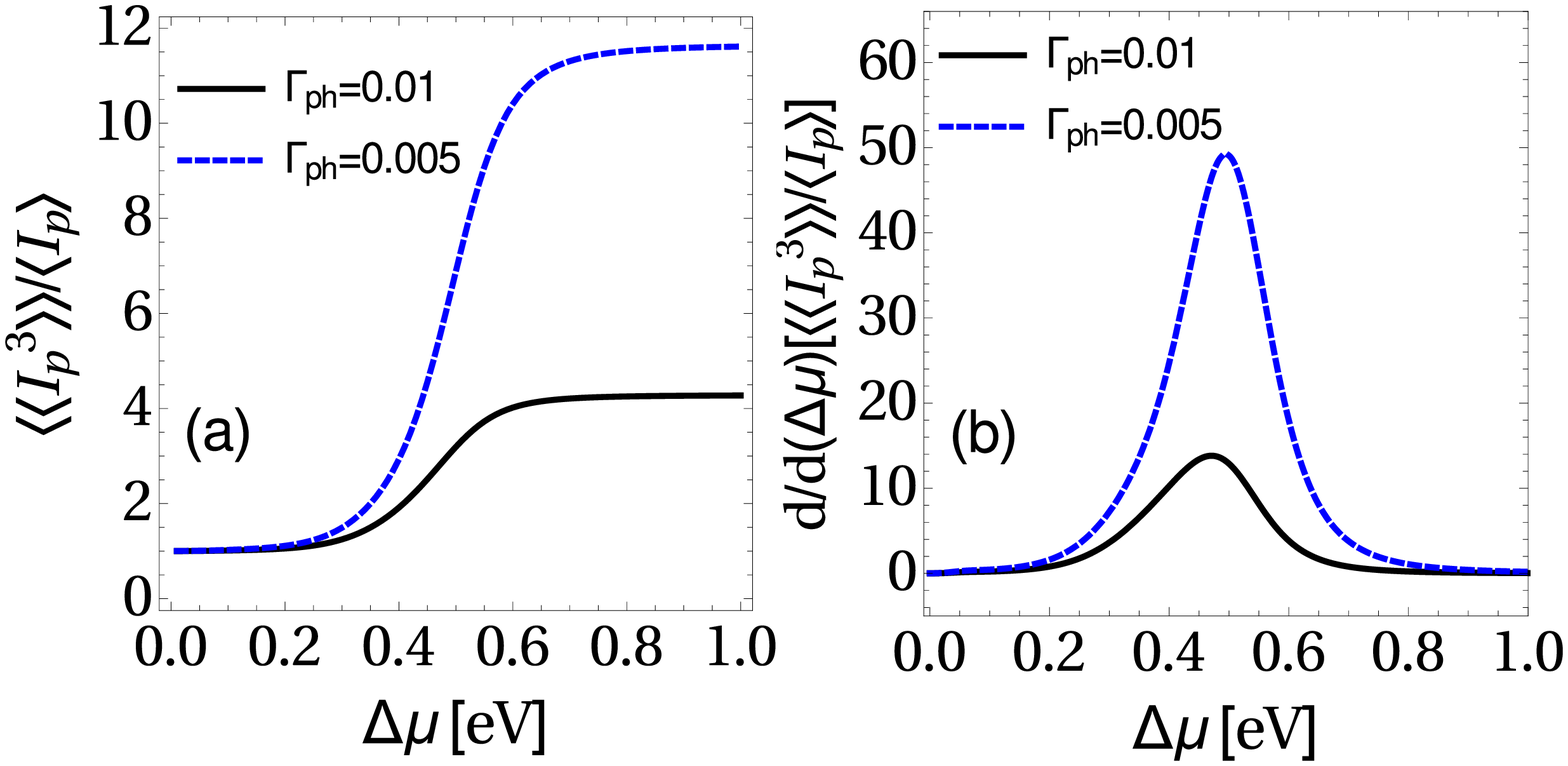}
\caption{(Color online)
Third cumulant of charge transfer as obtained from QME and NEGF (overlapping).
We used $\Gamma=0.1$ eV,
other parameters are same as in Fig. (\ref{mean-ph}).
}
\label{skew}
\end{figure}

In Fig. \ref{current} we plot the average current and the differential conductance, and observe similar trends as in the average phonon number.
This can be understood from Eq. (\ref{eq:avg-current-phonon}),
where the current is expressed in terms of the mean phonon number.
The current increases in two steps, and it saturates at high values of the bias. 

Figures \ref{fano} and \ref{skew} illustrate the Fano factor $F= \langle S_p\rangle /\langle I_p \rangle$
and the normalized skewness $ \langle I_p^3\rangle  /\langle I_p \rangle$, respectively.
We find that results obtained from QME calculation exactly match NEGF,
even beyond the weak-coupling regime, for $\Gamma=0.1$ eV.


\subsection{Scaling relations at high bias} 
\label{anal}

We derive here scaling relations for the mean phonon number and transport quantities,
current and high order cumulants, in the limit of low electronic temperature and high bias.
We further assume strong metal-molecule coupling and isolate
the primary mode from the secondary phonon bath, $\Gamma_{\rm ph}=0$.
In the wide band approximation, we can write down
the spectral function as
$J_{\alpha}(\omega)= 4g/\Gamma_{\alpha}$, $\alpha=L,R$.
We now introduce the notation $\bar g^2 = 8g^2/\pi$, assume
a zero electronic temperature and  apply a large bias, $\Delta\mu=\mu_R-\mu_L\gg \epsilon_{d,a},\omega_0$.
The rates (\ref{eq:kel}) reduce to
\bea
k_d&\approx&k_d^{R \to L}\approx \frac{\bar{g}^2 \omega_0} {\Gamma_L \Gamma_R} \Big(\frac{\Delta \mu}{\omega_0}+1\Big), \nonumber \\
k_u&\approx&k_u^{R \to L}\approx \frac{\bar{g}^2 \omega_0} {\Gamma_L \Gamma_R} \Big(\frac{\Delta \mu}{\omega_0}-1\Big).
\eea
In this limit, the average phonon number and the mean displacement for the oscillator are  given as
\be
\langle \hat n \rangle = \frac{1}{2} \Big(\frac{\Delta \mu}{\omega_0}-1\Big),
\quad \langle \hat X^2 \rangle = \frac{\Delta \mu}{\omega_0}. 
\ee
In analogy with the classical oscillator which satisfies equipartition (in terms of mean square displacement $=\frac{1}{2\omega_0} \langle X^2 \rangle$),
we define the effective temperature,
$T_{eff}= \Delta\mu/2$, which is solely determined by the voltage drop across the junction \cite{Dima}.
%

The cumulant generating function for charge current, can be simplified as well in this limit,
\be
{\cal G}(\lambda_p)= \frac{\bar{g}^2 \omega_0}{\Gamma_L \Gamma_R} \Big[1- \sqrt{1+ \Big(\frac{\Delta\mu^2}{\omega_0^2}-1\Big)\Big(1-e^{2 i \lambda_p}\Big)}\Big].
\ee
It yields the current and the noise as
\bea
\langle I_p \rangle &=& \frac{\bar{g}^2 \omega_0} {\Gamma_L \Gamma_R} \Big[\frac{\Delta\mu^2}{\omega_0^2} -1\Big], \nonumber \\
\langle S_p \rangle &= &\frac{\bar{g}^2 \omega_0}{\Gamma_L \Gamma_R} \Big[\frac{\Delta \mu^4}{\omega_0^4} -1\Big].
\eea
%
The Fano factor now reduces to $F= 1 + (\Delta\mu/\omega_0)^2$, showing a $\Delta\mu^2$ scaling \cite{Belzig}.
It can be further demonstrated  that leading-order nonlinear contributions of higher order cumulants
obey the scaling relation
\be
\frac{C_{n+1}}{C_n} \propto \Big(\frac{\Delta\mu}{\omega_0}\Big)^2,
\ee
with $C_n$ as the $n$-th order cumulant.
It is interesting to note the strong nonlinear dependence on bias in the present model.
similarly to that obtained with the single-site Anderson-Holstein problem
\cite{Utsumi1,Belzig,vanKampen}. 

\subsection{Charge rectification}
\label{Rect}

The proposal for an organic molecular rectifier (diode), based on donor-acceptor molecules,
had largely initiated the field of molecular electronics \cite{Ratner-Aviram}.
It is thus interesting to test the operation of a charge diode
in the present model system, to gather simple guidelines for optimizing this effect.

\begin{figure}
\centering
\includegraphics[width=0.45\textwidth]{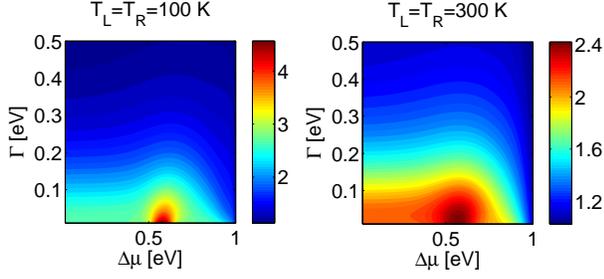}
\caption{
(Color online) Rectification ratio in the donor-acceptor junction
with $\epsilon_a=0.2$, $\epsilon_d=0$, $\omega_0=0.05$,
$g=0.01$, $\Gamma_{\rm ph}=0.005$ eV, and  $T_{\rm ph}=300$ K.
The temperature of the metal electrodes are
(a) $T_L=T_R=100$ K, and (b) $T_L=T_R=300$ K.
}
\label{rectif}
\end{figure}

It is obvious that to act as a diode, the junction as sketched in Fig. \ref{Fig1}
should posses a spatial asymmetry. This can be introduced
e.g. by using distinct D and A units, to support asymmetric states, $\epsilon_d\neq  \epsilon_a$.
The metal-molecule hybridization energy should be made small though, so as not to conceal this asymmetry
by level broadening.
This situation parallels with design principles for standard tunneling diodes \cite{vanderzant15,Ratnerpn15,Kilgour15}.
We apply the voltage in a symmetric manner, and for simplicity, we keep the levels fixed, independent of bias.
If $\epsilon_d<\epsilon_a$, the functionality of the system as a diode is expected to be optimized when $(\mu_R-\mu_L)/2\sim\epsilon_a$;
in the forward direction the two electronic levels are placed within the bias window beneficial for conduction, while
in the opposite bias $\epsilon_a$ is placed far away from $\mu_R$.
Fig. \ref{rectif} presents a contour map of the rectification ratio, defined as 
\bea
R\equiv\frac{\langle I_p\rangle_+}{|\langle I_p\rangle_-|},
\eea
with $+$ ($-$) identifying the application of
forward (reversed) bias, $\mu_R-\mu_L>0$ ($\mu_L-\mu_R>0$).
Increasing the temperature of the metals washes away the diode behavior, as a
electrons can cross the junction
even when the Fermi function is placed in a region of low density of states. Similarly,
reducing the frequency of the vibrational mode weakens the diode effect since even low-temperature
electrons can excite the mode.
%

\section{Conclusion}
\label{Summary}

We studied the non-equilibrium characteristics
of vibrationally-assisted electronic conduction in a two-site junction,
and demonstrated an equivalence between the QME and the NEGF approaches
beyond the standard {\it weak} metal-molecule coupling regime.
Examined non-equilibrium observables include the mean phonon number and
the charge current and its higher fluctuations.
Our study provided us with some important-general lessons:
(i) Perturbation expansions.
%
The master equation approach employed here
was developed while taking into account all electron scattering processes
with an exchange of a single quanta with the vibration.
In the language of the NEGF, an RPA scheme had to be employed for
including all these processes,
crucial for obtaining the correct steady state results. 

(ii) Vibrational coherences.
What role do vibrational coherences play in transport of electrons in our model,
at weak electron-phonon coupling?
While an NEGF-based derivation does not readily address this question, our QME method
was based on rate equations for the populations of the vibrational states,
decoupling the contribution of off-diagonal terms in the reduced density matrix.
The agreement between these tools
indicate that quantum coherences contribute to phonon occupation number and transport characteristics
only in higher orders of the electron-vibration coupling.

(iii) Full counting statistics.
NEGF theory provided us with a close analytical form for the CGF, [Eq. (\ref{eq:CGF-RPA})].
One should note that the CGF was achieved here given the harmonic nature of the primary mode and
its linear coupling---through its displacement---to the
different environments.
More complex models with e.g. anharmonic interactions, cannot be readily treated
with the NEGF. In contrast, the QME approach as described here offers a straightforward program for simulating
the current and its cumulants with general many-body molecular interactions.

Future studies will aim in reconciling QME and NEGF treatments in other models, e.g.,
when involving direct electron tunnelling between sites,
and in generalizing the present results to include high order terms (strong) in electron-phonon interactions.

\section*{Acknowledgments}
This work was funded by an NSERC Discovery Grant, the Canada Research Chair program,
and the CQIQC at the University of Toronto.


\renewcommand{\theequation}{A\arabic{equation}}
\setcounter{equation}{0}  
\section*{Appendix: Counting field dependent bath induced rates}

The harmonic mode is coupled to metal electrodes and to a phonon environment. These baths
induce transitions within the mode, and the rates are written in Sec. \ref{model}.
The current and its cumulants are reached by defining counting-fields dependent rates
for processes driven by the electron baths. We identify these rates by the tilde symbol,
\begin{widetext}
\bea
\big[\tilde{k}^{\rm el}_{d}\big]^{L \to R} &=& \int \frac{d\epsilon}{2\pi} f_L(\epsilon) (1-f_R(\epsilon+\omega_0)) J_L(\epsilon) J_R(\epsilon+\omega_0)
 e^{-i(\lambda_p + (\epsilon + \omega_0)\lambda_e)}, \nonumber \\
\big[\tilde{k}^{\rm el}_{d}\big]^{R \to L} &=&  \int \frac{d\epsilon}{2\pi} f_R(\epsilon) (1-f_L(\epsilon+\omega_0)) J_R(\epsilon) J_L(\epsilon+\omega_0)
 e^{i(\lambda_p+ \epsilon \lambda_e)},
\nonumber\\
\tilde{k}^{\rm el}_u&=&\tilde{k}^{\rm el}_d[\omega_0 \rightarrow -\omega_0].
\eea
\end{widetext}
We further define the full electronic rates as
\bea
 \tilde{k}^{\rm el}_{u/d} = \big[\tilde{k}^{\rm el}_{u/d}\big]^{L \to R} + \big[\tilde{k}^{\rm el}_{u/d}\big]^{R \to L},
\label{eq:klamel}
\eea
and the total transition rates affecting the primary mode,
\bea
\tilde{k}_{u/d}&=&\tilde{k}^{u/d}_{\rm el} + k^{\rm ph}_{u/d}.
\label{eq:klamtot}
\eea
Note that processes induced by the phonon bath do not include here counting fields.
We also define the difference between electronic rates as
\bea
\big(k^{\rm el}_{u}\big)'\equiv \partial_{i\lambda_p} \tilde{k}^{\rm el}_{u}|_{\lambda=0}
= \big[k^{\rm el}_{u}\big]^{R \to L} \!-\! \big[k^{\rm el}_{u}\big]^{L \to R}
\nonumber\\
\big(k^{\rm el}_{d}\big)'\equiv \partial_{i\lambda_p} \tilde{k}^{\rm el}_{d}|_{\lambda=0}
 = \big[k^{\rm el}_{d}\big]^{R \to L} \!-\! \big[k^{\rm el}_{d}\big]^{L \to R}.
\label{eq:klamdiff}
\eea
These expressions come to play in the Born-Markov treatment of the characteristic function,  Sec. \ref{CGF-QME}.

\end{document}